\documentclass[epjST]{svjour}

\usepackage{graphicx}
\usepackage{graphics}
\usepackage{bm}
\usepackage{epsfig}
\usepackage{amssymb}
\usepackage{amsmath}

\newcommand{\be}{\begin{equation}}
\newcommand{\ee}{\end{equation}}
\newcommand{\ba}{\begin{eqnarray}}
\newcommand{\ea}{\end{eqnarray}}
\newcommand{\bi}{\begin{itemize}}
\newcommand{\ei}{\end{itemize}}
\newcommand{\tr}{{\rm Tr\,}}

\newcommand{\nn}{\nonumber \\}

\newcommand{\half}{{\textstyle\frac{1}{2}}}

\newcommand{\<}{\langle}
\renewcommand{\>}{\rangle}
\newcommand{\eq}{Eq.~}

\newcommand{\fig}{Fig.~}

\newcommand{\la}{\label}

\newcommand{\txts}{\textstyle}

\newcommand{\ud}{\,\mathrm{d}}

\begin{document}
\title{Hadron Structure on the Lattice}
\author{Harvey B. Meyer\inst{1}\fnmsep\thanks{\email{meyerh@kph.uni-mainz.de}} }
\institute{Institut f\"ur Kernphysik, 
Johannes Gutenberg Universit\"at Mainz, 
55099 Mainz, Germany}
\abstract{
A few chosen nucleon properties are described
from a lattice QCD perspective: the nucleon sigma term and the scalar
strangeness in the nucleon; the vector form factors in the nucleon,
including the vector strangeness contribution, as well as parity
breaking effects like the anapole and electric dipole moment; and
finally the axial and tensor charges of the nucleon. The status of the
lattice calculations is presented and their potential impact on
phenomenology is discussed.
} 
\maketitle
\section{Introduction} \label{intro}

One of the goals of lattice QCD simulations is to calculate the
properties of hadrons from first principles and to understand how
their structure arises from QCD.  In this review we will focus on the
structure of the nucleon, because it is the only stable hadron in the
Standard Model and because a precise knowledge of its structure has
implications for new physics searches.

Many properties of the nucleon can be calculated on the lattice,
see~\cite{Hagler:2009ni} for a comprehensive review.  In addition one
has the option not available experimentally to vary the quark masses,
which can help to make contact with idealized models. This ability
represents one more handle to explore `the internal landscape of the
nucleon', to use the expression of the 2007 NSAC\footnote{The
  U.S. Nuclear Science Advisory Committee.} Long Range Plan.

It is useful to distinguish between those properties that are well
determined experimentally, and others where the lattice can
potentially be as accurate or more accurate than phenomenology in the
foreseeable future. The first class then serves as a set of control
quantities; these are for instance the electromagnetic form factors
and the axial charge of the proton.  The second class contains for
instance the sigma term, the vector strangeness form factor and the
isovector tensor charge, which are much harder to extract accurately
from experiments. However, in order to increase confidence in the
lattice predictions for the second class of observables, it is
necessary to make contact between the lattice postdictions of the
`control quantities' and their experimental values.

It should be noted that calculations in the mesonic sector are at a
more advanced stage, in the sense that several benchmark quantities
match the phenomenological determinations, and have in some cases even
overtaken them in accuracy. Examples thereof are $f_K/f_\pi$, form
factors of pseudoscalar mesons, and low-energy constants of chiral
perturbation theory~\cite{Sachrajda:2011tg,Colangelo:2010et}.

Rather than attempting to give a comprehensive overview of lattice
nucleon structure calculations, we discuss three topics: the nucleon
mass decomposition and the associated scalar matrix elements; the
standard electromagnetic form factors as well as the anapole and
electric dipole moments; and finally the axial and tensor charges.
In this way both mature calculations and exploratory studies 
will be covered. 

\section{The nucleon mass decomposition} \label{sec:decompo}

Let $T_{\mu\mu}$ be the trace of the energy-momentum tensor, 
\be
\la{eq:Tmumu} T_{\mu\mu} = {\txts\frac{\beta(g)}{2g}}
G^a_{\mu\nu}G^a_{\mu\nu} \;+\; (m_u\, \bar u u + m_d\, \bar d d) +
m_s\, \bar s s + \dots 
\ee 
with $\beta(g)=-b_0g^3+\dots$ and $b_0=(\frac{11}{3}N_c-\frac{2}{3}N_{\rm f})(4\pi)^{-2}$.
The expectation value of this operator on a
nucleon at rest yields the nucleon mass~\cite{Shifman:1978zn,Ji:1994av}, 
\be
\la{eq:TmumuN} \frac{\<N|\int d^3x \,T_{\mu\mu}|N\>}{\<N|N\>} = M_N.
\ee 
Each contributing operator in (\ref{eq:Tmumu}) is gauge invariant
and renormalization group invariant. It is of interest to calculate
the relative size of these contributions to \eq(\ref{eq:TmumuN}).
Recently, there have been several lattice calculations of the nucleon
sigma term, defined as 
\be  \la{eq:sigmaN1}
\sigma_N \equiv m_{ud} \<\bar u u + \bar d d\>,
\qquad m_{\rm ud}\equiv \half(m_u+m_d).
\ee 
The expectation value refers to the operator evaluated on the zero-momentum
nucleon state, with the vacuum expectation value subtracted.
In a mass-independent scheme, the quantity can also
be obtained from the Feynman-Hellmann theorem, 
\be 
\sigma_N = m_{\rm  q} \frac{\partial}{\partial m_{\rm q}} M_N .  
\ee
To leading order
in chiral perturbation theory, the derivative with respect to the
quark mass can be replaced by the derivative with respect to the pion
mass, 
\be \la{eq:sigmaN}
\sigma_N \simeq \half M_\pi \frac{\partial}{\partial
  M_{\pi}} M_N \equiv \tilde\sigma_N.  
\ee

\subsection{The nucleon sigma term and 
the quark mass dependence of the nucleon mass}\label{sec:MNmq}

In phenomenology, the light-quark scalar form factor can be related to
the $\pi N$ scattering amplitude at the Cheng-Dashen point
$t=+2m_\pi^2$~\cite{PhysRevLett.26.594}.  Correcting to get the scalar
matrix element at zero momentum transfer, the value obtained
is $\sigma_N  = 45$MeV~\cite{Gasser:1990ce}. 
More recent experimental data leads to larger values~\cite{Pavan:2001wz}.

\begin{figure}
\centerline{
\includegraphics[width=0.50\textwidth]{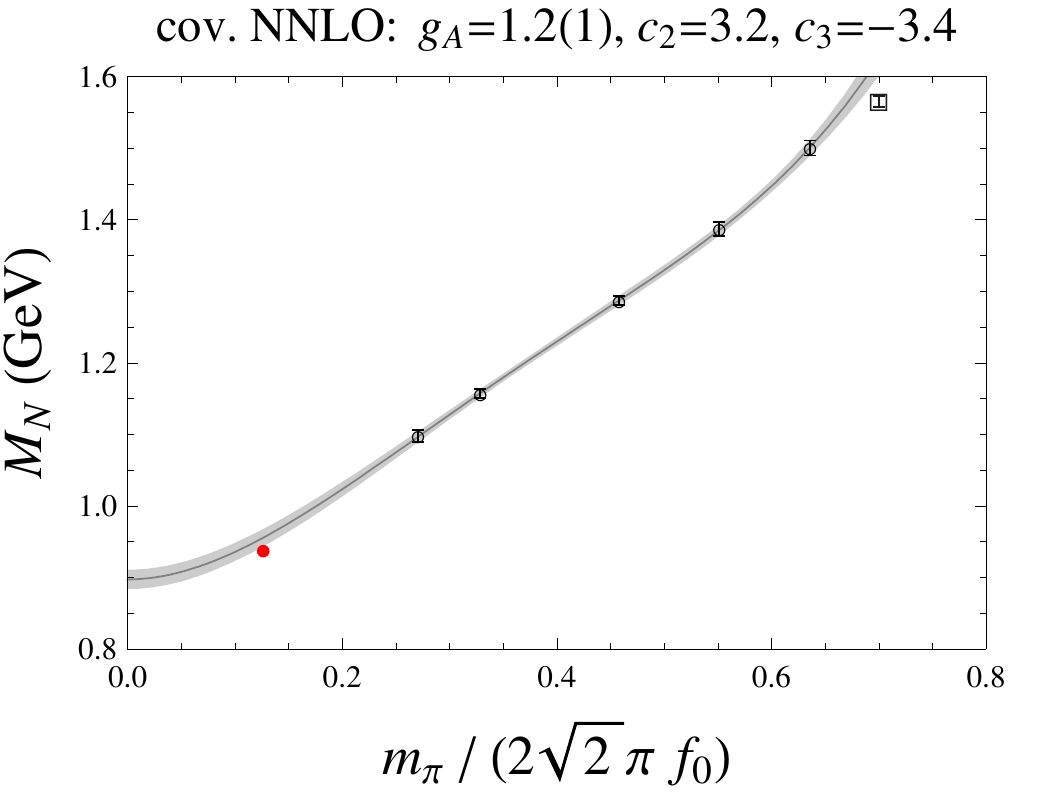}
            \includegraphics[width=0.50\textwidth]{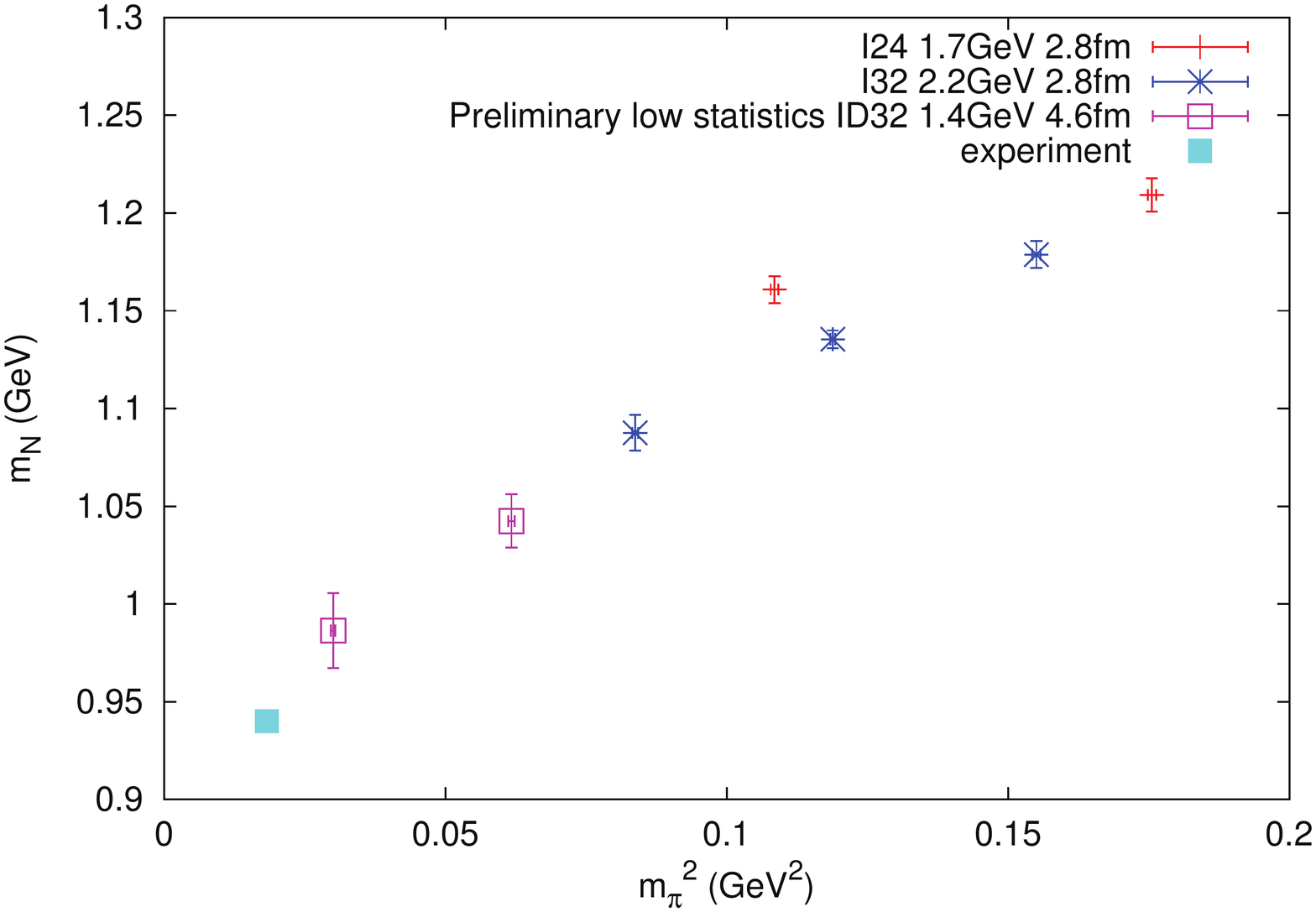}}
\centerline{
\includegraphics[width=0.50\textwidth]{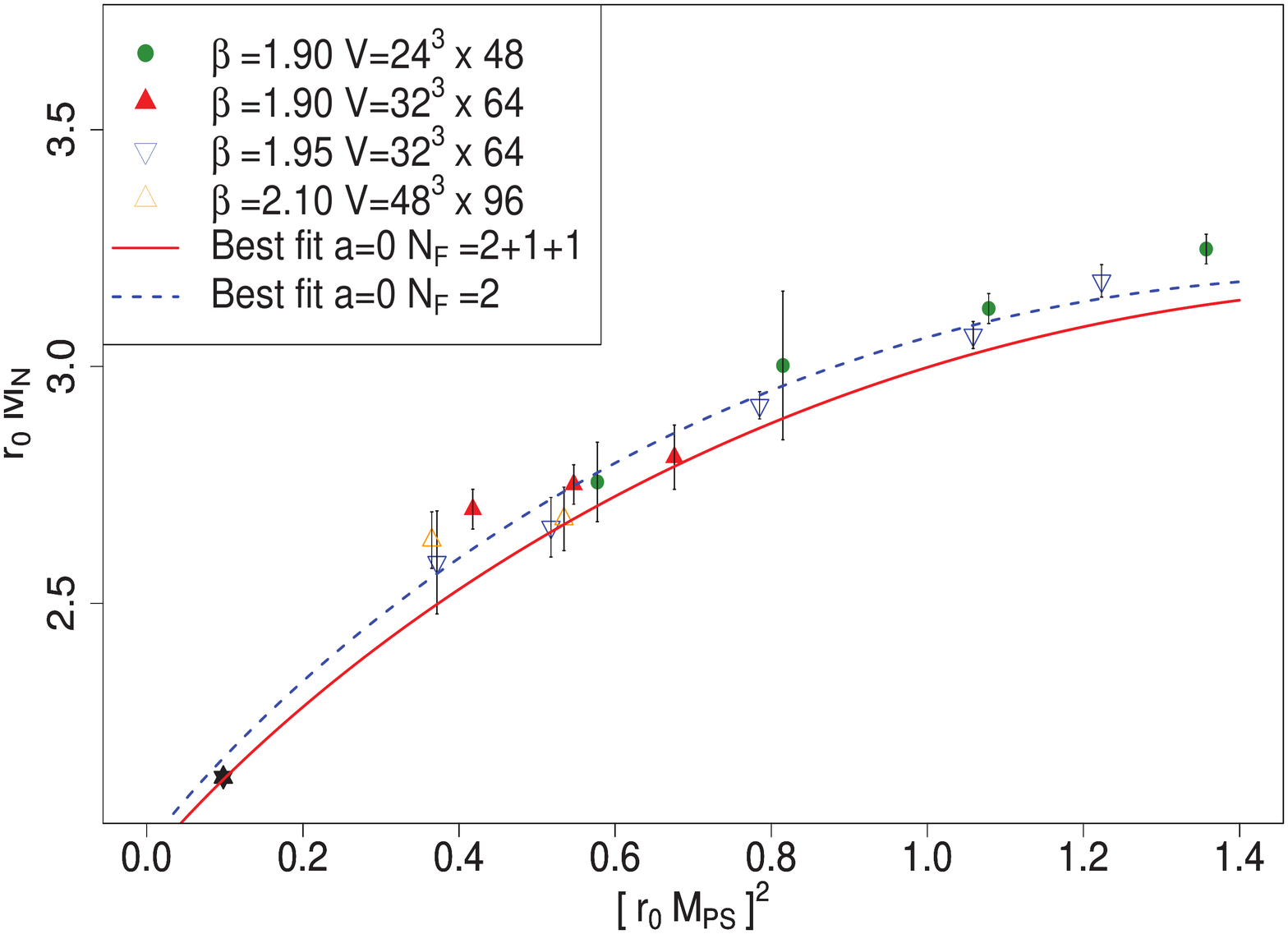}
            \includegraphics[width=0.50\textwidth]{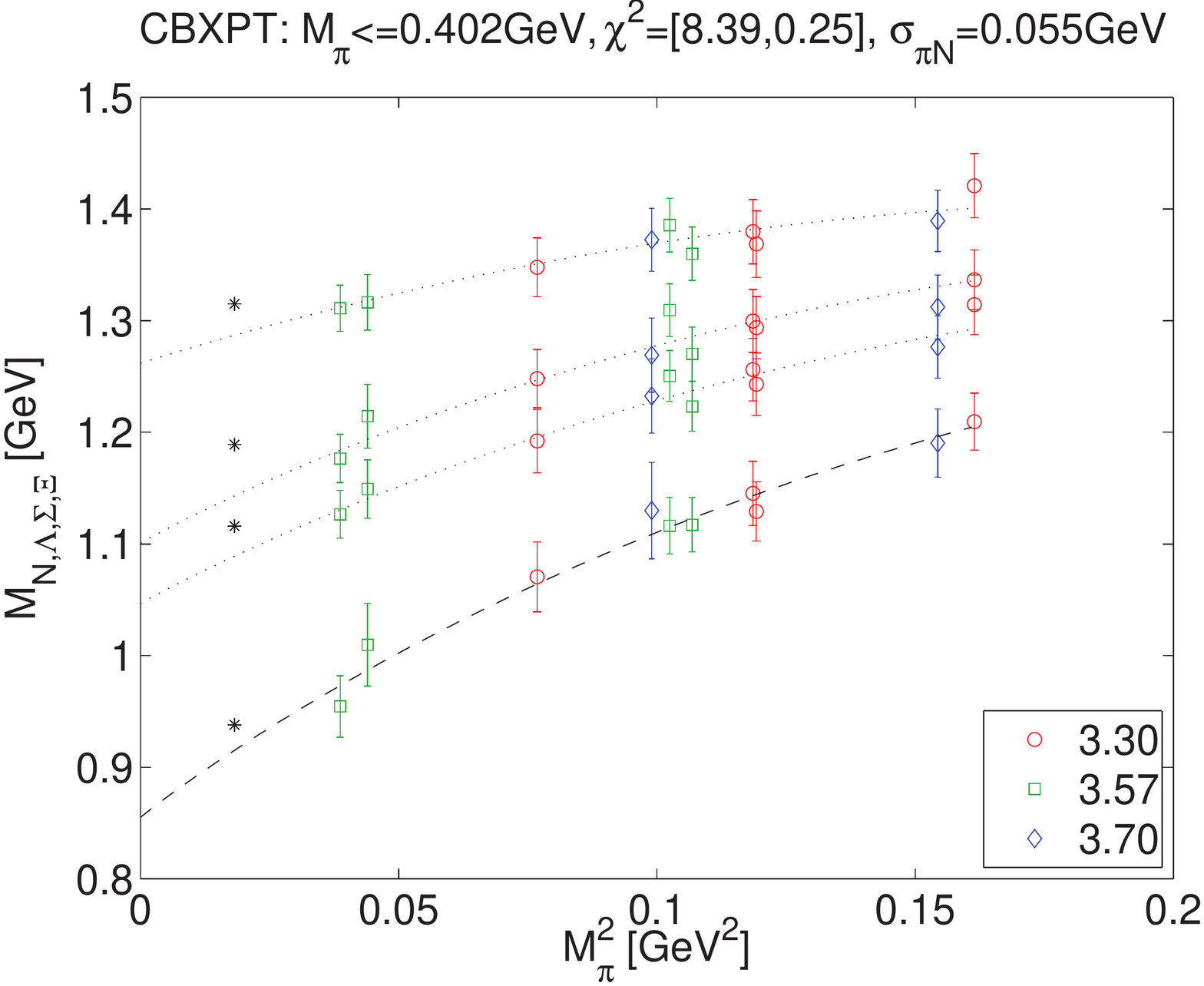}}
\caption{Recent lattice calculations by four different collaborations 
         of the pion mass dependence of the nucleon mass.
\underline{Top}: LHP (\cite{WalkerLoud:2008bp}, $N_{\rm f}=2+1$, 
                mixed action domain-wall/staggered)  and RBC-UKQCD collaboration
               (\cite{Ohta:2011nv}, $N_{\rm f}=2+1$ domain-wall fermions).
\underline{Bottom}: ETM (\cite{Drach:2010hy}, 
                  $N_{\rm f}=2+1+1$ twisted mass Wilson fermions) 
                and BMW collaborations (\cite{Durr:2010ni}, 
                  2+1 stout-smeared Wilson fermions).}
\label{fig:nucleon-mass}
\end{figure}

On the lattice, the nucleon mass has been calculated for a range of
pion masses.  This set of data points is fitted with a functional form
provided by a chiral effective theory. Via \eq(\ref{eq:sigmaN}) the
parameter $\tilde\sigma_N$ is one of the fit parameters. In this way,
the LHP collaboration extracted the value $\tilde\sigma_N \simeq
42(17)$MeV in an NNLO SU(2) covariant baryon ChPT formula without
explicit delta-baryon degrees of freedom~\cite{WalkerLoud:2008bp} (see
top left panel of \fig\ref{fig:nucleon-mass}). The lightest pion mass
reached is 295MeV.  When the $N-\Delta$ mass splitting is treated as
being small and the delta as being an active degree of freedom,
Walker-Loud et al. find $\tilde\sigma_N \simeq
84(27)$MeV~\cite{WalkerLoud:2008bp}. The latter fit is however poorly
constrained, due to the many parameters involved in the fit; for
instance, the nucleon-delta coupling was set to $c_A=1.5(3)$.

Young and Thomas~\cite{Young:2009zb} find $\tilde\sigma_N \simeq
47(10)$MeV by fitting LHP~\cite{WalkerLoud:2008bp} and
PACS-CS~\cite{Aoki:2008sm} octet baryon spectrum data using a finite range
regularization ansatz~\cite{Young:2002ib}.  The ETM collaboration
obtains $67(8)$MeV in a two-flavor calculation with pion masses down to
about 300MeV~\cite{Alexandrou:2008tn}. The JLQCD collaboration finds
$\tilde\sigma_N =
52(2)(^{+20}_{-7})(^{+5}_{-0})$MeV \cite{Ohki:2008ff} in
$N_{\rm f}=2$ QCD, where the first uncertainty is statistical, the
second comes from the chiral extrapolation and the third is an
estimate of finite-volume effects.  The central value comes from
fitting the pion mass dependence of the nucleon mass by a fourth-order
polynomial in $m_\pi$, where the O($m_\pi$) vanishes and the
coefficient of the $m_\pi^3$ term is known in terms of the axial
charge $g_A$ and the pion decay constant $f_\pi$.  The BMW
collaboration recently presented the result $\tilde\sigma_N \simeq
55(10)_{\rm stat}$MeV~\cite{Durr:2010ni} using an ansatz based on a
modified integration contour in covariant baryon ChPT that exploits
the approximate SU(3) flavor symmetry. The bottom right panel of
\fig(\ref{fig:nucleon-mass}) displays such a fit to the octet of
baryons with 7 parameters and 40 data points. Different fit ans\"atze
and ranges are estimated to lead to a systematic uncertainty of about
10MeV~\cite{Durr:2010ni}.

There is thus a satisfactory agreement among lattice calculations, as
well as between $\tilde\sigma_N$ calculated on the lattice and
$\sigma_N$ obtained from experimental pion-nucleon scattering.  The
uncertainty on the lattice results to date are comparable to the
phenomenological uncertainty and the two kinds of determinations are
in good agreement. Of course it is desirable to improve the accuracy
of these determinations, not least in view of the importance of this
matrix element in dark matter searches~\cite{Ellis:2008hf}.

\subsection{Strangeness in the nucleon} \label{sec:strange}

The size of the strangeness term in \eq(\ref{eq:Tmumu}) is often
parametrized by
\be
y \equiv \frac{2\<\bar ss\>}{\<\bar u u + \bar d d\>},
\ee
where the expectation value has the same meaning as in \eq(\ref{eq:sigmaN1}).

In phenomenology, a standard way to estimate $y$ is to extract the
matrix element $\sigma_8 \equiv m_{\rm ud} \<\bar u u + \bar d d - 2
\bar s s\>$ from the octet baryon spectrum. A benchmark ChPT estimate
is $\sigma_8 \simeq 36(7)$MeV~\cite{Gasser:1982ap,Borasoy:1996bx}.
The difference between $\sigma_{N}$ and $\sigma_8$ is then attributed
to the strange quarks, $\sigma_8 = \sigma_{N} (1-y)$.  Gasser et
al. thus estimated $y=0.2$ in 1991~\cite{Gasser:1990ce}.  The more
recent estimates from the $\pi N$ scattering data~\cite{Pavan:2001wz}
lead to a large value for $y$, 0.3--0.6. Such as large value is
surprising from the quark model point of view.

In view of the large value and the large uncertainty on the
phenomenological estimates of $y$, it is of interest to study this
quantity ab initio using lattice computational techniques.  
Young and Thomas~\cite{Young:2009zb} found $\frac{\sigma_s}{M_N}=0.033(16)(4)(2)$
at the same time as $\sigma_N$ by fitting the baryon octet spectrum
(as discussed in section \ref{sec:MNmq}).
The calculations we review below have been done 
by evaluating directly the forward matrix element of $\bar s s$.
We start with the recent results of the JLQCD collaboration~\cite{Takeda:2010cw}. 
It employs overlap fermions, which preserve a
lattice form of chiral symmetry exactly~\cite{Luscher:1998pqa}.  On
$N_{\rm f}=2$ ensembles (i.e. with a quenched strange quark), the
JLQCD collaboration finds
\be
\frac{\sigma_s}{M_N} 
\equiv \frac{m_s}{M_N} \<N|\bar s s|N\>
 = 0.032(8)_{\rm stat}(22)_{\rm syst}.
\ee
Takeda of the JLQCD collaboration has also presented preliminary results 
for the same quantity calculated on $N_{\rm f}=2+1$ ensembles 
of overlap fermions~\cite{Takeda:2010id},
\be
\frac{\sigma_s}{M_N} =0.013(12)(16).
\ee
Toussaint and Freedman~\cite{Toussaint:2009pz} on the other hand find 
\be
\frac{\sigma_s}{M_N}=0.063(6)(9)
\ee
using 2+1 flavors of Kogut-Susskind fermions.

\begin{figure}
\centerline{\includegraphics[width=0.82\textwidth]{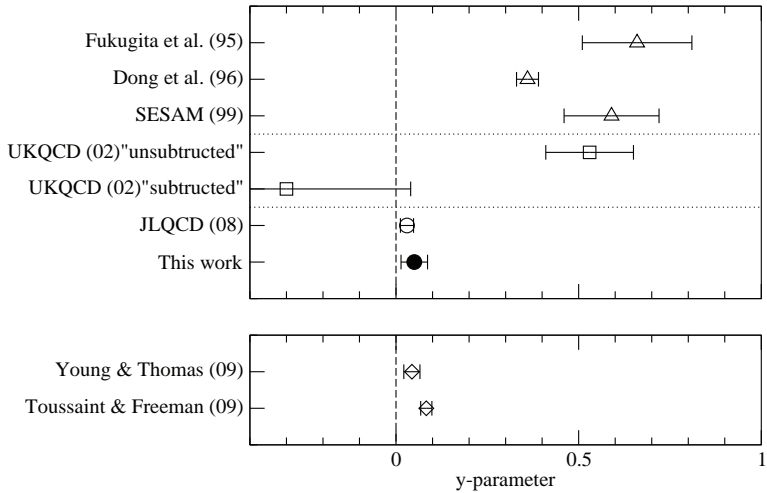}}
\caption{Summary of lattice $y$ parameter calculations from~\cite{Takeda:2010cw}.}
\label{fig:strangeness}
\end{figure}

Thus while there are still noticeable differences between the results,
there is a consensus among these recent calculations that
$\frac{\sigma_s}{M_N}<0.08$, and possibly it is even much smaller.  To
put this into perspective, we note that in 2+1 flavor QCD, the result
for a very massive strange quark would be $\sigma_s/M_N \simeq 2/29\simeq
0.069$~\cite{Shifman:1978zn,Kryjevski:2003mh}. At the physical value
of the strange quark mass, the results reviewed above appears to be
somewhat smaller.  Finally we remark that the successive contributions
of the charm, bottom and top quarks to the mass sum rule are not
expected to decrease with the mass of the quarks, they are predicted
in perturbation theory to be about 0.086, plus/minus
$5\%$~\cite{Kryjevski:2003mh}.

These results on the scalar strangeness can also be translated into
predictions for the $y$ parameter, if the quark mass ratio $m_s/m_{\rm
  ud}$ is known. Figure (\ref{fig:strangeness}) is reproduced
from~\cite{Takeda:2010cw}, where this mass ratio was set to the value
27.4 for the conversion and the value for $\sigma_N$ taken
from~\cite{Ohki:2008ff}. The recent results lead us to the conclusion
that $y<0.1$, unlike some of the older calculations, which were
affected by an uncontrolled operator mixing problem due to the lack of
chiral symmetry on the lattice~\cite{Michael:2001bv}. They also
suggest that the aforementioned phenomenological determinations of $y$
overestimate this parameter.


\begin{figure}
\centerline{\includegraphics[width=0.60\textwidth]{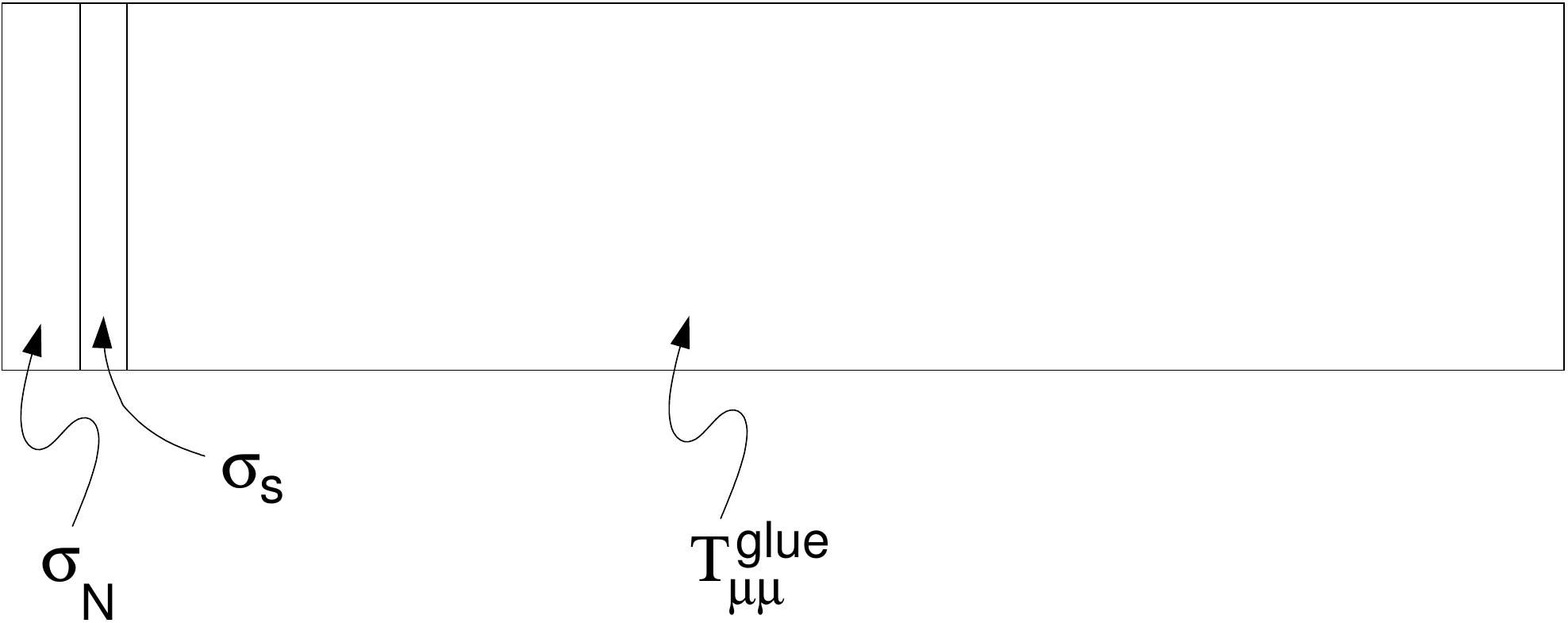}}
\caption{Decomposition of the nucleon mass in $N_{\rm f}=2+=1$ QCD 
based on \eq(\ref{eq:Tmumu}) according to the 
lattice QCD results of the JLQCD collaboration~\cite{Ohki:2008ff,Takeda:2010cw}.}
\label{fig:partit}       
\end{figure}

\subsection{Summary}\label{sec:summary}

The relative size of the light-quark, strange-quark and glue
contributions to the nucleon mass is illustrated in
\fig(\ref{fig:partit}).  The central values used in this figure are
those of the JLQCD collaboration~\cite{Ohki:2008ff,Takeda:2010cw}.
The relative uncertainty on $\sigma_N$ and $\sigma_s$ is still large.  It is
interesting that the light quarks and the strange quark make
contributions of the same order, both of them being small compared to
the gluonic contribution. This weak dependence on the quark mass is
the qualitative behavior expected in the heavy-quark regime, and
appears to extend down into the light-quark regime.

It is worth noting that in the 1990's, the quark mass contribution 
to the nucleon mass was estimated to be quite large. For instance,
Ji put forward the numbers~\cite{Ji:1994av}
\[
\frac{1}{M_N} \big(\sigma_N + \sigma_s \big) \simeq 0.11,~~~ 0.17,
\]
where the two values come from treating the strange quark mass 
as heavy and light, respectively. The JLQCD result,
with the strange quark partially quenched~\cite{Ohki:2008ff,Takeda:2010cw},
is smaller, $0.09(3)$.

\section{Electromagnetic form factors 
and the anapole and dielectric moments} \label{sec:EMFF}

In this section we discuss the matrix elements of the vector current
between two nucleon states. We start with the standard case where no
parity violating effects are present (such as the the $\theta$ angle
or the weak force). First, we review the form factors of the isovector
current, which is technically easier to calculate on the lattice.
Then we present some recent results on the strangeness form factors.
We then move on to discuss the anapole moment and the electric dipole 
moment of the nucleon, reviewing the few existing calculations and 
commenting on the prospects of future calculations.

The matrix elements of the vector current between two nucleon states
(with no source of parity violation) reads
\ba
 \<p',s'|J^\mu|p,s\> = \bar u_{s'}(p') \Gamma^\mu(q^2)u_s(p)\,, 
\\
\Gamma^\mu(q^2) = \gamma^\mu {F_1}(q^2) 
               + i\sigma^{\mu\nu}\frac{q_\nu}{2M_N}{F_2}(q^2)
\ea
with $\sigma^{\mu\nu} = \frac{i}{2}[\gamma^\mu,\gamma^\nu]$.
Furthermore we will show results for the Sachs form factors, 
\ba
G_E(Q^2) &=& F_1(Q^2)  - \frac{Q^2}{(2M_N)^2} F_2(Q^2),
\\
G_M(Q^2) &=& F_1(Q^2) + F_2(Q^2).
\ea

\subsection{Isovector contributions} 

\begin{figure}
\centerline{\includegraphics[width=0.45\textwidth,angle=0]{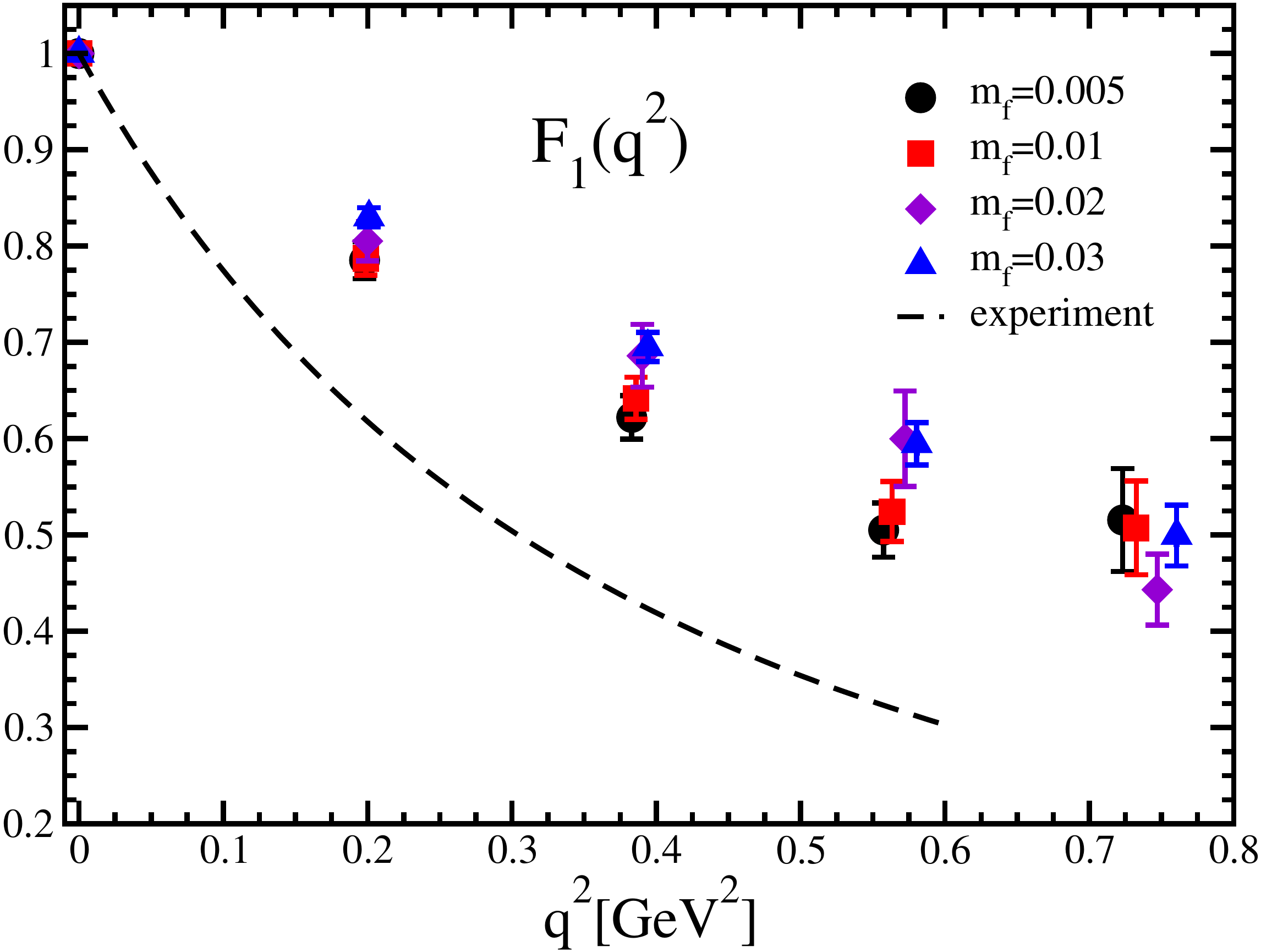}
         \includegraphics[width=0.48\textwidth,angle=0]{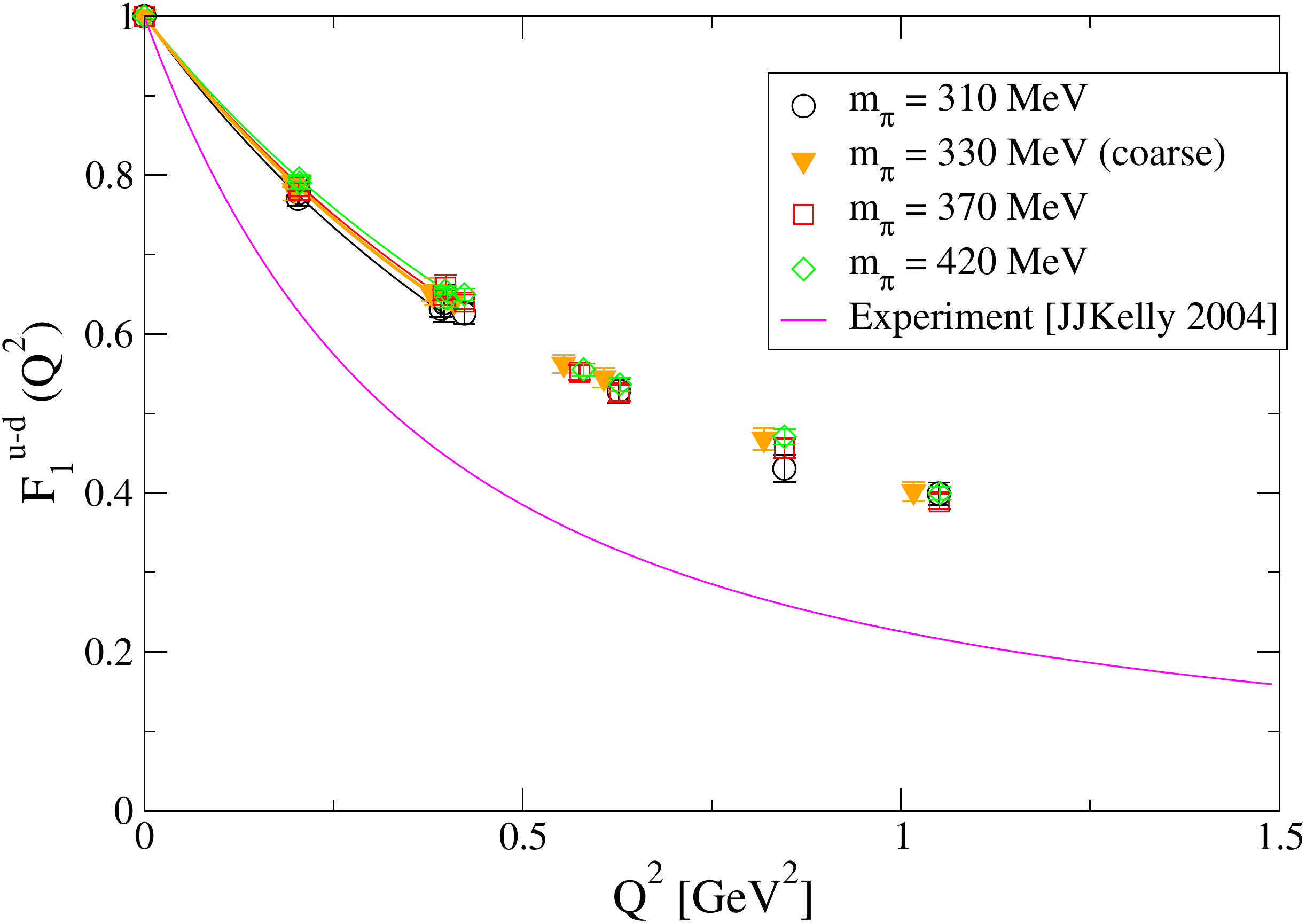}}
\caption{The isovector Dirac form factor $F_1^{u-d}$. \underline{Left}: 
$N_f=2+1$ domain-wall-fermion calculation by the RBC-UKQCD collaboration
at a lattice spacing $a=0.114$fm and for $m_\pi\geq 350$MeV~\cite{Yamazaki:2009zq}.
\underline{Right}: calculation by Syritsyn et al. (LHP collaboration~\cite{Syritsyn:2009mx})
at a lattice spacing $a=0.084$fm and for $m_\pi\geq 310$MeV on configurations generated 
by the RBC-UKQCD collaboration~\cite{Allton:2008pn}. The lattice data is compared to 
Kelly's parametrization~\cite{Kelly:2004hm}.}
\label{fig:F1}       
\end{figure}

Calculating the nucleon form factors of the vector isospin current
$J_\mu^a$ is an important task for lattice QCD. The disconnected Wick
contraction diagrams cancel for degenerate $u,d$ quarks, which
simplifies the calculation considerably. Ultimately the goal is to
make contact with the experimental measurements, and unless this
contact is made, the degree of confidence one will have in lattice
calculations of nucleon structure will be limited.  A number of
collaborations have carried out calculations of the isovector form
factors for pion masses down to about 280MeV. A few calculations exist
at lighter pion masses, but they remain exploratory, either because of
the lack of control of finite-volume effects, or because the
statistical fluctuations on the nucleon correlator increases
drastically.

As an example, \fig(\ref{fig:F1}) displays the isovector Dirac form
factor calculated with 2+1 flavors of domain-wall
fermions~\cite{Yamazaki:2009zq,Syritsyn:2009mx}.  Our main
observations are the following.  A high level of statistical accuracy
has been achieved in the displayed range of pion masses. Secondly the
pion mass dependence of the form factor is very weak between 500 and
300MeV, and in this range the form factor falls off much more slowly
in $Q^2$ than the experimentally measured form factor. Thirdly, the
dipole form $1/(1+Q^2/M_D^2)^2$ provides a good fit up to $Q^2\approx
1$GeV$^2$.  The pion mass dependence of the dipole mass $M_D$ is of
physical interest.  It it natural to normalize it by the corresponding
nucleon mass, however this does not resolve the difficulty of
extrapolating the lattice data points~\cite{Meyer:2010tx}.  The dipole
mass can also be normalized by the $\rho$ meson mass computed at the
same quark mass.  This ratio turns out to be significantly larger than
it is in the real world~\cite{Meyer:2010tx}, about 1.5 instead of 1.1.

\begin{figure}
\centerline{\includegraphics[width=0.75\textwidth,angle=0]{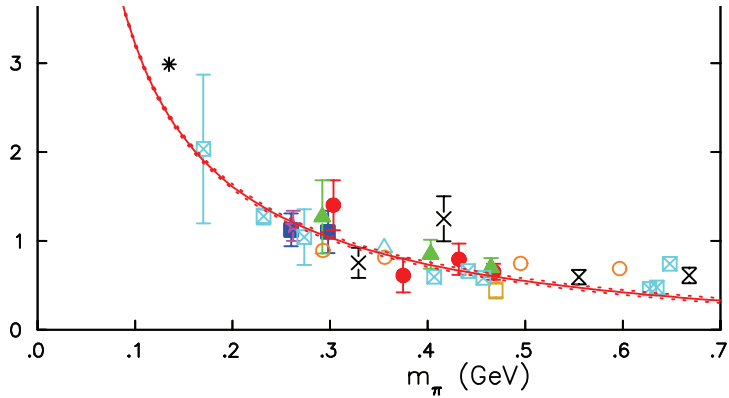}}
\vspace{-1cm}
\caption{The chiral extrapolation of the combination $(r_2)^2 \kappa_v$ in units
of $({\rm fm}^2 \mu_N)$,
Fig. from~\cite{Alexandrou:2011db}.  Data from $N_{\rm f} = 2$ twisted
mass fermions~\cite{Alexandrou:2010hf} ($a = 0.089$fm: filled red
circles for $L = 2.1$fm and filled blue squares for $L = 2.8$fm; $a =
0.070$fm: filled green triangles for $L = 2.2$fm; $a = 0.056$fm: purple
star for $L = 2.7$fm and open yellow square for $L = 1.8$fm); $N_{\rm f}
= 2 + 1$ domain-wall fermions with $a = 0.114$fm and $L =
2.7$fm~\cite{Yamazaki:2009zq} (crosses); mixed action
calculation~\cite{Bratt:2010jn} (open orange circles for $L = 2.5$fm
and open cyan triangles for $L = 3.5$fm); and $N_{\rm f}=2$ O($a$)
improved Wilson fermions~\cite{Pleiter:2011gw} (cyan cross-in-square).
}
\label{fig:chiralfit}
\end{figure}

The small $Q^2$ region of the form factors is parametrized in leading
order by the anomalous magnetic moment as well as the Dirac and Pauli
radii, for instance
$-6\frac{dF_1}{Q^2}|_{Q^2=0}=(r_1)^2=\frac{12}{M_D^2}$.  At present
these have to be extrapolated to the physical pion mass.  At
sufficiently small pion masses, the functional form is predicted by
chiral effective theory.  The asymptotic chiral behavior of the Dirac
and Pauli radii as well as of the anomalous magnetic moment are 
\ba
(r_1^v)^2&\sim& \log m_\pi\,, 
\\ 
(r_2^v)^2&\sim& m_\pi^{-1}\,,
\\ 
\kappa_v\equiv F_2(0) &\sim& {\rm cst}\,.  
\ea 
More precisely, as an example of
the expressions involved, we give here~\cite{Bernard:1998gv} 
\be
\la{eq:combi} \kappa_v(m_\pi)(r_2^v)^2 = \frac{g_A^2M_N}{8\pi f_\pi^2
  m_\pi} + \frac{c_A^2 M_N}{9\pi^2f_\pi^2\sqrt{\Delta^2-m_\pi^2}}
\log\left[{\txts\frac{\Delta}{m_\pi}}+\sqrt{{\txts\frac{\Delta^2}{m_\pi^2}}-1}\right],
\ee 
where $f_\pi\simeq 86$MeV in the chiral limit, $\Delta\simeq 293$MeV
is the nucleon-delta mass splitting and $c_A$ is the axial
nucleon-delta coupling. In this `small-scale expansion' power counting
scheme, the delta resonance is an `active' degree of freedom.  The
combination (\ref{eq:combi}) has the advantage that one low-energy
constant drops out.  The extrapolation of this expression is
illustrated in \fig(\ref{fig:chiralfit})~\cite{Alexandrou:2011db}.
The data of several collaborations appear on the figure. The
coefficient of the $1/m_\pi$ term is expressed in terms of well-known
quantities, and the lattice data and the experimental data point
cannot be joined by the ansatz (\ref{eq:combi}). The validity of the
fit ansatz is presumably limited to pion masses much smaller than
those at which accurate lattice data is currently available. An ansatz
to tame the strong pion mass dependence has been
proposed~\cite{Wang:2008vb}, where the form at higher quark masses is
inspired by quark models. However lattice data at smaller pion masses
will clearly be required for a controlled calculation of the proton radii 
and anomalous magnetic moment.


\subsection{Strangeness vector form factor: Wick-disconnected contributions}

The number of calculations of the strange quark contribution to the
electromagnetic form factor is much more limited. At a numerical level
they are important as a technical step towards calculating the
Wick-disconnected diagram contributions to the $u,d$ form factors. At
a physics level they teach us about the spatial distribution of
strange quark-antiquark pairs in the nucleon.

The Wick contraction of the three-point function that yields the
strange form factor is purely disconnected.  This means that the
expectation value of the strange vector current on nucleon states is
mediated entirely by gluons. As a consequence, the Monte-Carlo
variance associated with the operator diverges as $1/a^6$, where $a$
is the lattice spacing.  This is the main reason Wick-disconnected
diagrams are difficult to calculate.  An additional difficulty is that
in order to select two of the initial- and final-state momenta $P$,
$P'$ and $Q=P'-P$ exactly on the lattice, the strange quark propagator
in the background gauge field would have to be computed for every
point in space, which is prohibitively expensive.  This difficulty is
cicumvented by obtaining the propagators stochastically.  This is
advantageous because the noise associated with this procedure can be
reduced below the noise level associated with the fluctuations of the
gauge fields at a cost which is still moderate compared to the cost of
generating the gauge fields.

\begin{figure}
\centerline{\includegraphics[width=0.35\textwidth,angle=-90]{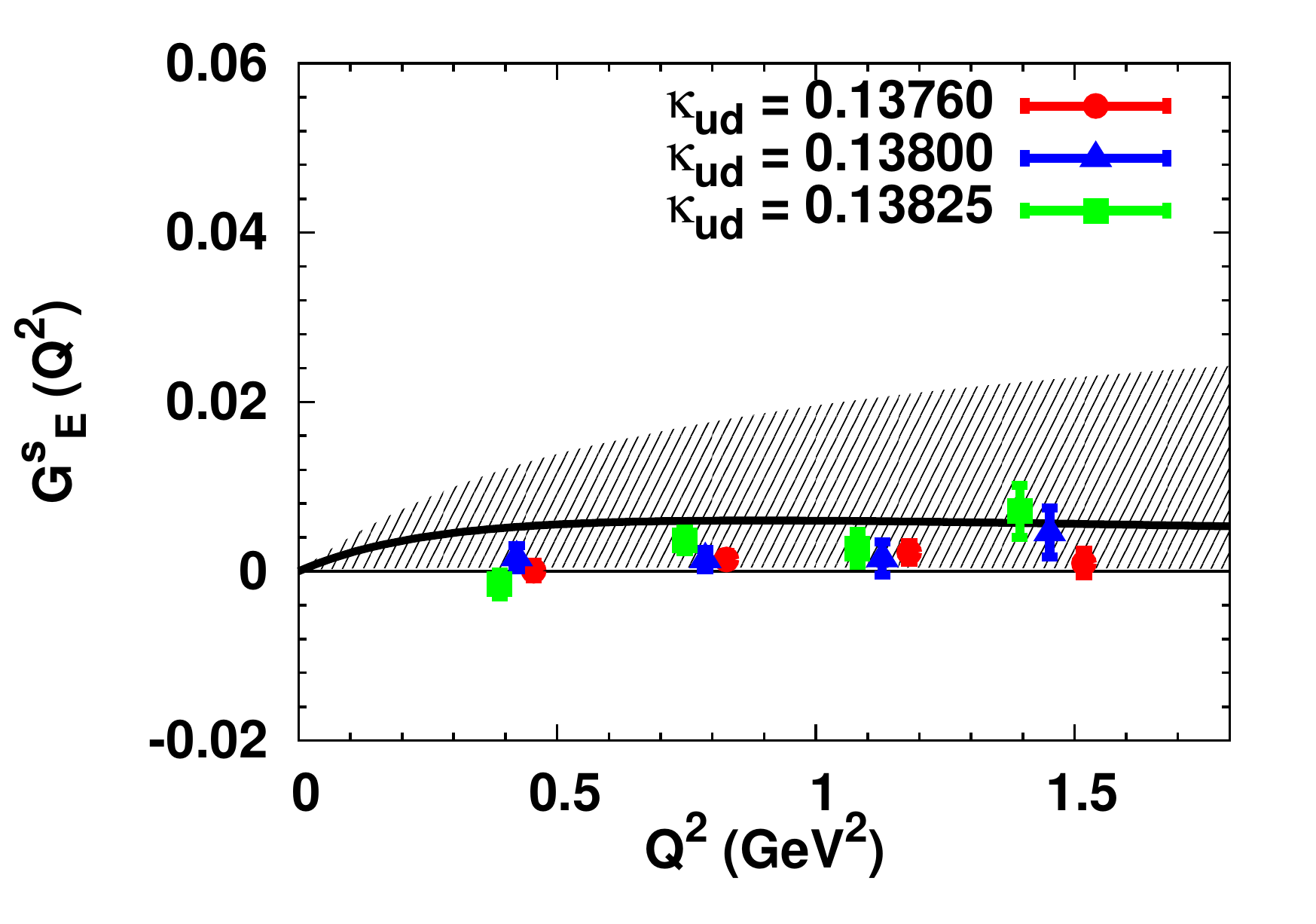}
         \includegraphics[width=0.35\textwidth,angle=-90]{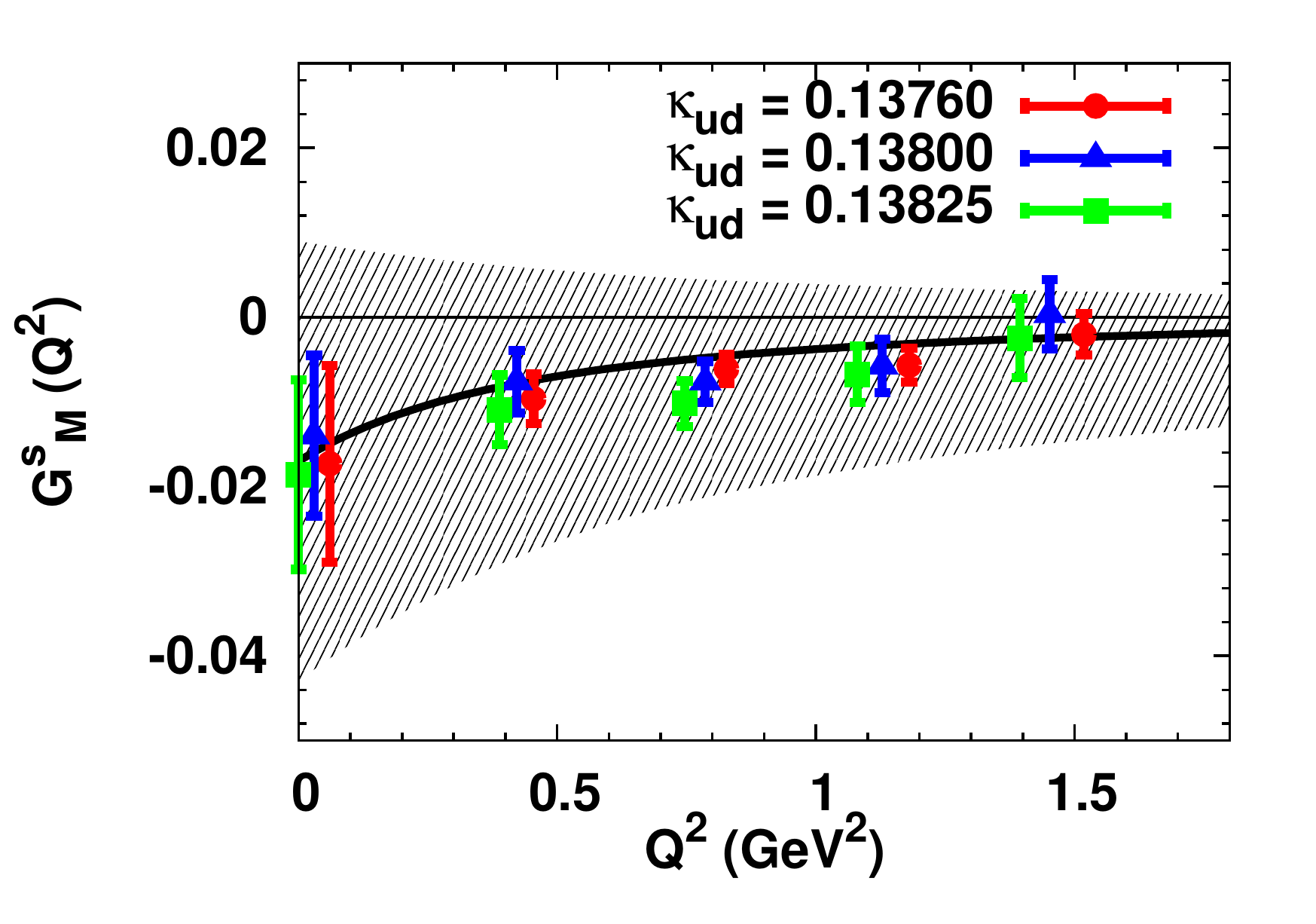}}
\caption{Strangeness Sachs form factors calculated at $m_\pi\geq 600$MeV
and at a lattice spacing $a=0.12$fm in $N_{\rm f}=2+1$ QCD~\cite{Doi:2009sq}.}
\label{fig:strang_vec}       
\end{figure}

\begin{figure}
\centerline{\includegraphics[width=0.75\textwidth,angle=0]{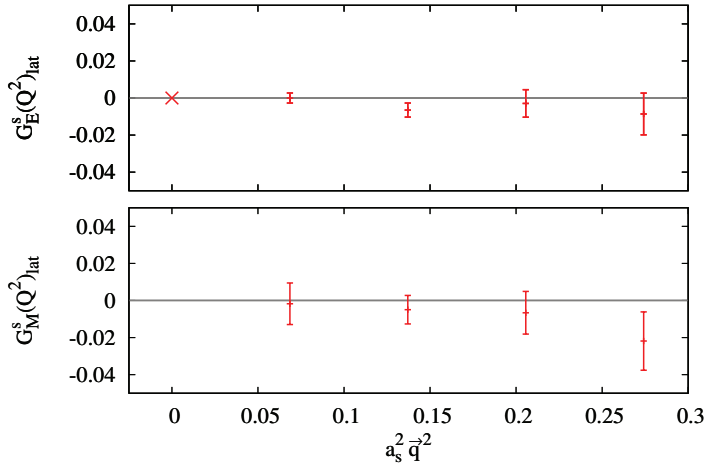}}
\caption{Strangeness Sachs form factors calculated on an anisotropic lattice with a spatial 
lattice spacing of $ a_s=0.108(7)$fm  and at $m_\pi=416(36)$MeV~\cite{Babich:2010at}.}
\label{fig:strang_vec2}
\end{figure}

The calculation of the $\chi$QCD collaboration is displayed in
\fig(\ref{fig:strang_vec}, Ref.~\cite{Doi:2009sq}).  It is restricted
to the regime of heavy pion masses $m_\pi\geq 600$MeV, but even
allowing for a conservative error band for the chiral extrapolation,
the results suggest that both the electric and magnetic form factors
are very small.  This is qualitatively confirmed by a very recent
calculation by Babich et al.  (\fig\ref{fig:strang_vec2},
Ref.~\cite{Babich:2010at}) at the lower pion mass of 416MeV. There is
some indication in both calculations that $G_M^{s}$ is negative.

Altogether these lattice results indicate that the strangeness form
factors of the nucleon are very small, even smaller than the bounds
obtained by recent experiments.  For comparison, the PVA4 experiment
at MAMI quotes $G_M^{s}(0.22{\rm GeV}^2) = -0.14\pm 0.11\pm
0.11$~\cite{Baunack:2009gy}. It will have to be seen whether the
strangeness form factors remain as small when the pion mass approaches
its physical value.

\subsection{Anapole form factor}\label{sec:anap}

There are several contexts in which parity violating effects in
nucleon structure measurements are important. One example is the
measurement of a neutron electric dipole
moment~\cite{Peng:2008ha}. Another is the parity violating scattering
experiments which aim at measuring the weak charge of the proton or
its strangeness form factor~\cite{Beck:2001yx}. 

If parity is not a good quantum number, 
two additional form factors are necessary to parametrize
the matrix elements of the vector current on a spin 1/2 bound state,
\ba
\Gamma^\mu(q^2) = \gamma^\mu {F_1}(q^2) 
+ i\sigma^{\mu\nu}\frac{q_\nu}{2M_N}{F_2}(q^2) 
+\left(\gamma^\mu\gamma_5 q^2 -2M_N \gamma_5 q^\mu\right){F_A}(q^2)
\\
+\sigma_{\mu\nu}\gamma_5  \frac{q_\nu }{2M_N} {F_3}(q^2).
\nonumber
\ea
The quantity $F_A(0)$ measures the anapole moment of the nucleon.  It
describes the fact that a matrix element of the vector current can
yield a result which has an axial-vector tensor structure.  Sources of
a non-zero $F_A$ form factor are the weak force or a potential
$\theta$ term in QCD.  Although we are not aware of any lattice
calculations of $F_A$, the contribution induced by a small $\theta$
term can be computed along the same lines as the electric dipole form
factor $F_3$ calculation described in the next section.



\subsection{Electric dipole moment}\label{sec:EDM}

The electric dipole moment is determined by the form factor $F_3$ at 
vanishing momentum tranfer,
\be
d_N=\frac{F_3(0)}{2M_N}.
\ee
When the only source of parity violation is the $\theta$-term of QCD,
the electric dipole moment has an expansion
\be
d_N=d_N^{(1)}\theta+{\rm O}(\theta^3),
\ee
and the goal of lattice simulations if to calculate $d_N^{(1)}$.
Together with experimental bounds on $d_N$, this quantity allows one
to derive an upper bound on $\theta$. Currently, the experimental
bound is $|d_N|<2.9\cdot 10^{-26}\,e\cdot{\rm  cm}$~\cite{Baker:2006ts}, 
and experiments are being planned to reach
an accuracy of (a few)$\cdot10^{-28}\,e\cdot{\rm  cm}$ 
(see~\cite{Peng:2008ha} for a review of these experiments).
The Standard Model contribution to the neutron EDM is less than 
$10^{-31}\,e\cdot{\rm  cm}$~\cite{Dar:2000tn}.

One method to obtain $d_N^{(1)}$ is to calculate the form factor
$F_3$, and extrapolate it to $Q^2=0$. As far as the Wick-connected
diagrams are concerned, smaller momentum transfers can be reached by
using twisted boundary conditions~\cite{Aoki:2008gv}.  There is an
alternative method. In a constant and uniform electric field ${\bf E}$, 
a shift in energy of the nucleon state with spin ${\bf S}$ takes place,
\be
\Delta E = d_N\, {\bf S\cdot E} + \dots
\ee
The dots refer to terms quadratic in the electric field (sensitive to polarizabilities)
and higher.

There are several methods to simulate QCD at small $\theta$. 
One is to explicitly Taylor-expand the observable around $\theta=0$, 
\ba
\<{\cal O}\>_\theta &=& \frac{1}{Z(\theta)}
\int  {\cal D }A_\mu {\cal D}\bar\psi{\cal D}\psi\, {\cal O}\,
e^{-S[A,\psi,\bar\psi]-i\theta\int d^4x \frac{g^2}{32\pi^2}\tr[G(x)\tilde G(x)]}
\\
&\simeq& \<{\cal O}\>_0 - i\theta \<Q{\cal O}\>_0.
\nonumber
\ea
In QCD  at $\theta=0$, the topological charge fluctuates around zero 
with a second moment
\be
\frac{\<Q^2\>}{V} = \frac{f^2m_\pi^2}{8} + \dots
\ee
where we have indicated the leading chiral behavior of $\<Q^2\>$
with $f\approx 130$MeV~\cite{Leutwyler:1992yt} and $V$ denotes the four-volume.

The other method involves simulating QCD at imaginary $\theta$, where 
no sign problem occurs. More precisely, the angle $\theta$ is rotated into a mass term,
\[ 
S_F = \bar\psi\,\{ D+\bar m + i(\bar\theta/N_{\rm f})\gamma_5 \bar m\}\,\psi\,,\qquad
\bar m = \cos(\bar\theta/N_{\rm f}), \qquad \bar\theta=N_{\rm f}\,\tan(\theta/N_{\rm f}).
\]
Then one sets $\bar\theta \doteq -i\,\bar\theta^{\rm \,I}$, $\theta^{\rm \,I}$ real.
The situation is similar to the case of a baryon chemical potential.

A further important aspect of the calculation is to take into account the
change in the polarization tensor of a nucleon propagator when the theta angle 
is switched on (see the discussion in~\cite{Shintani:2005xg,Berruto:2005hg}).

\begin{table}
\begin{center}
\begin{tabular}{|l|l|l|}
\hline
   &    $F_3(q^2)$   &  $\Delta E = d_N \, {\bf S\cdot E}$     \\
\hline
Taylor &
\begin{minipage}{0.32\textwidth}\vspace{0.1cm}
{ Shintani et al ($N_{\rm f}=2\!+\!1$)~\cite{Shintani:2005xg};
\\{Blum et al~\cite{Berruto:2005hg}}}\end{minipage} &
\begin{minipage}{0.34\textwidth}{ Shintani et al ($N_{\rm f}=2\!+\!1$)~\cite{Shintani:2006xr};\\
 Shintani et al~\cite{Shintani:2008nt} }\end{minipage} \\
\hline
$i\theta$ &   {\footnotesize {Horsley et al~\cite{Aoki:2008gv}}} &      \\
\hline
\end{tabular}
\end{center}
\caption{Different lattice methods to calculate $d_N^{(1)}$. The calculation of Horsley et al. used 
twisted boundary conditions on the fermion fields. When not indicated, the flavor content 
is $N_{\rm f}=2$.}
\label{tab:theta}
\end{table}

The available results in the $N_{\rm f}=2$ theory, 
which should be regarded as preliminary, are summarized as follows,
\bi
\item Blum et al. 2005~\cite{Berruto:2005hg}: $d^{(1)}_N\lesssim 20\cdot 10^{-3}\,e$ fm;
\item QCDSF 2008~\cite{Aoki:2008gv}: ~~~$d^{(1)}_N\lesssim 50\cdot 10^{-3}\,e$ fm.
\ei
The results were obtained at a quark mass only slightly smaller than the physical strange quark mass.
The results obtained directly on the lattice thus provide a relatively loose bound
compared to the magnitude of the pion loop contribution to $d^{(1)}_N$
(Crewther et al.~\cite{Crewther:1979pi}), 
\be
d^{(1)}_N \approx 3.6\cdot 10^{-3} \, e\cdot{\rm fm}.
\ee
In the future, when an actual value for $d_N^{(1)}$ is obtained, 
the chiral extrapolation of $d_N^{(1)}$ will however be strongly constrained by 
the fact that it vanishes in the chiral limit~\cite{Crewther:1979pi},
\be
d^{(1)}_N \sim m_\pi^2 \log m_\pi^2.
\ee
Therefore calculating $d_N^{(1)}$ on the lattice remains a challenging
but realistic and important goal.

\section{Axial and tensor charge of the nucleon}\label{sec:axial}

The isovector axial charge is the forward matrix element of the axial current
on a polarized nucleon,
\be
\<P,S|A^a_\mu |P,S\> = \bar U(P,S)\gamma_\mu\gamma_5 \frac{\tau^a}{2}U(P,S)\cdot g_A.
\ee
It is related to helicity parton distribution functions by the Bjorken sum rule
\be
g_A =  \int_0^1 \ud x \,\big(\Delta q(x) + \Delta\bar q(x) \big). 
\ee

We focus here on the isovector combination, which is measured in
neutron $\beta$ decay with the result $g_A^{\rm expt}=1.2695(29)$ in
units of the vector coupling $g_V$.  The isovector axial charge
$g_A\equiv g_A^{3}$ does not have a renormalization scale dependence,
although the higher moments of $\Delta q(x)$ do.

In several respects, $g_A$ is a cornerstone of nucleon physics.  The
Adler-Weisberger sum rule relates the departure of $g_A$ from unity to
an integral over the difference between the $\pi^{+}p$ and $\pi^{-} p$
scattering cross sections, where the the delta features as a prominent
resonance. Furthermore in the chiral limit the axial charge is
directly related to the pion-nucleon coupling strength by the
Goldberger-Treiman relation $g_A = f_{\pi} g_{\pi NN}/M_N$.

\begin{figure}
\centerline{\includegraphics[width=1.0\textwidth]{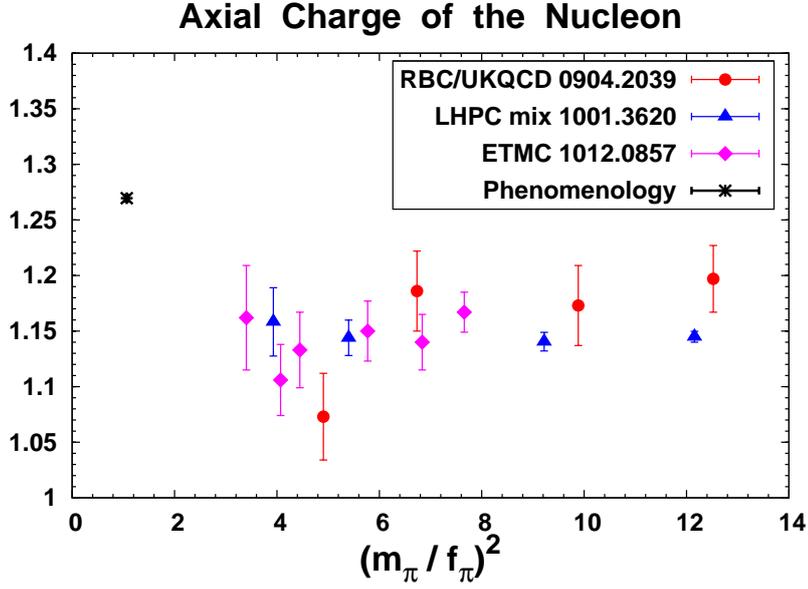}}
\caption{Summary of several recent lattice $g_A$ calculations.}
\label{fig:gA}
\end{figure}

A summary of recent published lattice calculations is given in
\fig(\ref{fig:gA}).  The results are very weakly pion-mass dependent,
and lie at a level of 1.15, i.e. about $10\%$ below the
phenomenological value. The displayed statistical errors are
significantly smaller than that, and it therefore remains to be seen
whether and how contact is made with the phenomenological value.  A
discussion of the sources of systematic error is given by H.~Wittig in
these same proceedings.

The chiral expansion of $g_A$ has been carried out to higher
order. The delta resonance is thought to influence the pion mass
dependence of $g_A$, since the gap between the nucleon and the delta
is about equal to the pion mass when
$m_\pi=350$MeV~\cite{WalkerLoud:2008bp}. In the small-scale expansion
counting scheme, the expression takes the form~\cite{Hemmert:2003cb}
\ba
 g_A(m_\pi)&=& g_A-\frac{g_A^3m_\pi^2}{16\pi^2 f_\pi^2}
~+~ 
4m_\pi^2\Big\{C(\lambda)+\frac{c_A^2}{4\pi^2f_\pi^2}
[{\txts\frac{155}{972}}g_1-{\txts\frac{17}{36}}g_A]+\gamma\log\frac{m_\pi}{\lambda}\Big\}
\nn
&&+~\frac{4c_A^2g_A}{27\pi f_\pi^2\Delta}m_\pi^3 
~+~ \frac{8c_A^2g_Am_\pi^2}{27\pi^2f_\pi^2}[1-{\txts\frac{m_\pi^2}{\Delta^2}}]^{\frac{1}{2}}\log R
\\
&& +~\frac{c_A^2\Delta^2}{81\pi^2f_\pi^2}\big(25g_1 - 57g_A\big)
\Big\{\log{\txts\frac{2\Delta}{m_\pi}}-[1-{\txts\frac{m_\pi^2}{\Delta^2}}]^{\frac{1}{2}}\log R\Big\},
\nn
\gamma&=&\frac{1}{16\pi^2f_\pi^2}\Big[\frac{50}{81}c_A^2g_1-\frac{1}{2}g_A-\frac{2}{9}c_A^2g_A-g_A^3\Big],
\\
R&=&\frac{\Delta}{m_\pi}+\Big[\frac{\Delta^2}{m_\pi^2}-1 \Big]^{\frac{1}{2}}.
\ea
The couplings $g_1$ and $c_A$ are respectively the axial delta-delta
and the axial nucleon-delta coupling, while $\Delta$ is the nucleon-delta
mass splitting in the chiral limit. In addition, when the $\Delta$ baryon 
is below the $N\pi$ threshold, as is the case in most lattice calculations to date,
$\sqrt{\Delta^2-m_\pi^2}\log R(m_\pi)$ is substituted by 
$-\sqrt{m_\pi^2-\Delta^2}\arccos(\Delta/m_\pi) $.
As usual, it is not a priori known what the effective radius of convergence is.
We remark that the $m_\pi^2\log m_\pi$ term has a negative coefficient
(most easily seen by setting $c_A=g_1=0$), 
which means that the axial coupling in the chiral limit is reached 
from above. This implies that the axial coupling must have a local maximum 
as a function of $m_\pi^2$ if we imagine a curve going through the lattice
data points and the phenomenological value. Such a structure remains to be seen
explicitly in the lattice data.

The other flavor-octet linear combination of axial charges, $g_A^8$
(proportional to the linear combination $(u+d-2s)$, where the SU(3)
generators are normalized according to $\tr\{\lambda^a\lambda^b\} =
2\delta^{ab}$) is of course also of interest, particularly in the
context of the quark spin contributions to the nucleon
spin~\cite{Jaffe:1987sx}. On the lattice, in addition to
Wick-connected diagrams, it requires calculating the difference of the
light-quark and strange-quark disconnected diagrams. The latter
however cancel at an $SU(3)_f$ symmetric point. The connected diagrams
at such a point were calculated for instance by the LHP collaboration
with all three quark masses equal to the physical strange quark
mass. In that case the result is
$g_A^{8}/g_A^{3}=0.315(9)$~\cite{Hagler:2007xi}.  The value in this
fairly massive theory is surprisingly close to the value extracted
from phenomenological octet baryon axial charges~\cite{Praszalowicz:2001ir}.


\subsection{The tensor charge}

\begin{figure}
\centerline{\includegraphics[width=0.86\textwidth]{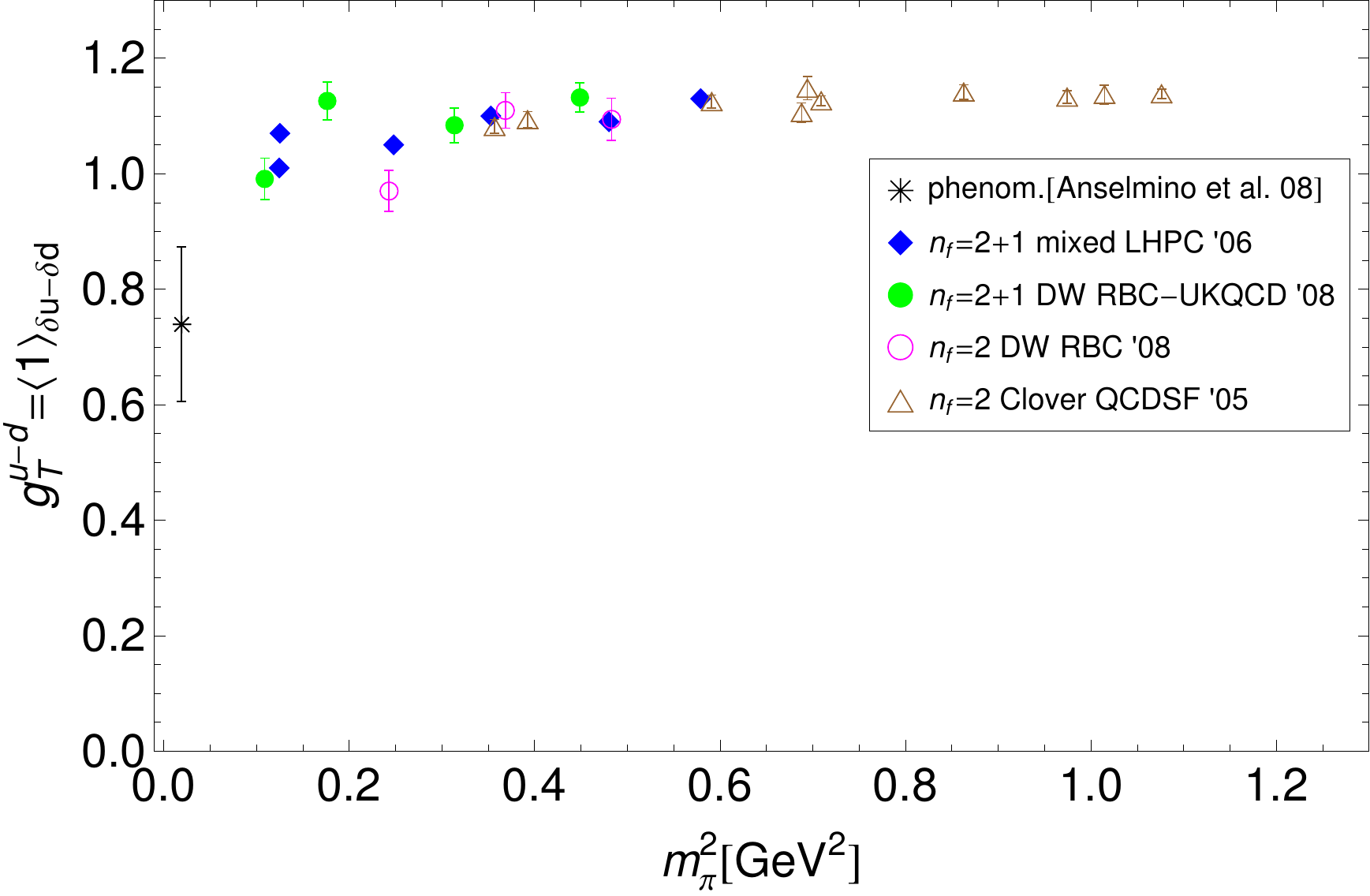}}
\caption{Summary of several recent lattice $g_T$ calculations~\cite{Hagler:2009ni}.
The phenomenological estimate is from~\cite{Anselmino:2008sj}. 
The references for the lattice calculations are in the same order as the caption
\cite{Edwards:2006qx}, \cite{Ohta:2008kd}, \cite{Lin:2008uz}, \cite{Gockeler:2005cj}.}
\label{fig:gT}
\end{figure}

Transversity is an active topic of research in deep inelastic
scattering and related experiments, see~\cite{Burkardt:2008jw} for a
review.
Similar to the axial charge for the longitudinal polarization, 
the tensor charge is the forward value of the antisymmetric-tensor form factor
\be
\<P,S| \bar q i\sigma_{\alpha\beta} q|P,S\>
= \bar U(P,S) i\sigma_{\alpha\beta} U(P,S) \cdot g_T.
\ee
In terms of the transversity parton distribution functions, it corresponds
to 
\be
g_T  = \int_0^1 \ud x \,  \big(\delta q(x) - \delta\bar q(x) \big),
\ee
i.e. it measures the $x$-average of the transverse polarization of
quarks (minus that of the antiquarks) in a transversely polarized,
fast moving proton. The fact that quarks and antiquarks appear with
opposite signs is physically significant. Here too we focus on the
isovector combination. Unlike the axial charge, the tensor charge has
a renormalization scale dependence. Figure (\ref{fig:gT}) displays
lattice results renormalized at an $\overline{\rm MS}$ scale of
$2$GeV.  Here the situation is opposite to the axial charge, in that
the lattice data points at $m_\pi\gtrsim 280$MeV are more accurate
than the phenomenological estimate. Also, the lattice data show the
right trend to approach the current phenomenological value of the
tensor charge. It is intriguing that the lattice data points for $g_A$
and $g_T$ are very close in value at a given pion mass (an observation already made
in~\cite{Khan:2004vw}), since in a non-relativistic 
theory they would be equal. The phenomenological values of $g_A$ and $g_T$ 
on the other hand are clearly split on either side of unity.

\section{Conclusion}\label{sec:concl}

Lattice calculations of nucleon structure are a vibrant area of
research which complements the worldwide experimental efforts
dedicated to unravel the structure of the nucleon. For some
quantities, I expect that in the coming decade a higher precision will
be achieved on the lattice than in experiments, for instance for the
isovector tensor charge (see \fig\ref{fig:gT}) or the strangeness
vector form factors (Figs.~\ref{fig:strang_vec},
\ref{fig:strang_vec2}). For hyperons, the same computational
techniques can be applied without any major additional
difficulties~\cite{Lin:2007ap}, while in an experiment hyperons
require a very different treatment.

With the development and improvement of techniques to handle
Wick-disconnected diagrams (see for
instance~\cite{Foley:2005ac,Bali:2009hu}), as they appear for instance
in strangeness matrix elements, the range of quantities that can be
studied is expanding significantly. At the same time, the results for
some of the more `prosaic' quantities such as the axial charge $g_A$
or the electromagnetic form factors $F^{u-d}_1$, $F^{u-d}_2$
absolutely need to be improved to the point where contact is
convincingly made with the well-established experimental
measurements. Once that is achieved the interplay of hadron structure
experiments and lattice calculations could be extremely fruitful.

\vspace{0.3cm}

I would like to thank Michael Ostrick for organizing a very
stimulating Symposium on `Many-body structure of strongly interacting
systems' and more widely my colleagues at the Institute of Nuclear
Physics in Mainz for a friendly work environment.

\bibliographystyle{JHEP}
\bibliography{/home/meyerh/CTPHOPPER/ctphopper-home/BIBLIO/viscobib}

\end{document}